\documentclass[12pt,preprint]{aastex}
\received{}
\accepted{}

\begin{document}

\title{\bf PLASMA EJECTIONS FROM A LIGHT BRIDGE IN A SUNSPOT UMBRA}
\author{AYUMI A. ASAI \altaffilmark{1,2}, 
TAKAKO T. ISHII \altaffilmark{1}, 
and HIROKI KUROKAWA \altaffilmark{1}}
\altaffiltext{1}{
Kwasan and Hida Observatories, Kyoto University, Yamashina-ku, 
Kyoto 607-8471, JAPAN}
\altaffiltext{2}{
Department of Astronomy, Kyoto University, Sakyo-ku, 
Kyoto 606-8502, JAPAN}
\email{asai@kwasan.kyoto-u.ac.jp}

\begin{abstract}

We present conspicuous activities of plasma ejections along 
a light bridge of a stable and mature sunspot in NOAA 8971 
on 2000 May 02. 
We found the ejections both in the H$\alpha$ (10$^4$~K) images obtained 
with the Domeless Solar Telescope (DST) at Hida Observatory, 
and in the 171 {\AA} (Fe {\sc ix}/{\sc x}, $\sim $10$^6$~K) 
images obtained with the Transition Region and Coronal Explorer ({\it TRACE}).
Main characteristics of the ejections are as follows: 
(1)Ejections occur intermittently and recurrently,  
(2)The velocities and the timings of the 171{\AA}-ejections are the same as 
those of H$\alpha$ ejections,
(3)The appearance of ejections are different from one another, 
i.e. the H$\alpha$ ejections have jet-like appearance, while that of 
171{\AA}-ejections is like a loop.

\end{abstract}


\section{INTRODUCTION}

Characteristics of H$\alpha$ surges have been studied for 
many years (e.g. Roy 1973). 
Those are summarized by \citet{Bru1977} as follows: 
(1)H$\alpha$ surges are straight or slightly curved spikes 
that are shot out of a small mound, 
(2)On the solar disk, they appear usually in absorption, 
but in their initial phase, sometimes in emission, 
(3)They also show strong tendency to recur. 
\citet{Kuro1988} and \citet{Kuro1993} have reported that 
H$\alpha$ surges are often found at the earliest stage of Emerging 
Flux Regions (EFRs) and continue recurrently for many hours.
They have also suggested that magnetic reconnection 
between a newly emerging flux and a pre-existing magnetic field 
is the essential mechanism of H$\alpha$ surge production.
\citet{Yok1995} showed in their numerical simulation 
that such a reconnection really produces H$\alpha$ surges in EFRs.

\citet{Roy1973} found that H$\alpha$ surges are also ejected from 
a light bridge of a sunspot umbra.
However, no detailed study has been made of plasma ejections from 
a light bridge until now.
We use the term ``light bridge'' to refer to a bright, long, and narrow 
feature penetrating or crossing a sunspot umbra. 
Light bridges are often seen in umbrae of mature and stable sunspots, 
and have been considered to have the same magnetic polarity as that of 
the sunspot umbrae, while their field strength is much weaker 
\citep{Bec1969}.

We found that conspicuous H$\alpha$ surge activities occurred along the 
light bridge of the sunspot umbra in the active region NOAA 8971 on 
2000 May 02, 
with 60 cm Domeless Solar Telescope (DST) at Hida Observatory, 
Kyoto University.
H$\alpha$ surge activities continued intermittently for about 6.5 hours, 
as long as the time span of our observation.
Examining the extreme-ultraviolet (EUV) images obtained 
with the Transition Region and Coronal Explorer ({\it TRACE}) 
\citep{Han1999,Sch1999}, 
we found similar ejections from the light bridge in 171 {\AA} 
(Fe {\sc ix}/{\sc x}) images.
Such ejections from a light bridge have never been reported before 
in the EUV wavelength.
From now on, we refer to the ejection seen in 171 {\AA} images 
obtained with {\it TRACE} as ``171{\AA}-ejection''. 

In this {\it Letter}, we report the morphological and the dynamical 
characteristics of the ejections from the light bridge, 
using the H$\alpha$ and the 171 {\AA} images of high spatial and 
temporal resolution obtained with DST and {\it TRACE}, respectively.
In \S 2, we summarize the observational data, 
and in \S 3, we present observational results and discussion. 
In \S 3.1, we report the features of the H$\alpha$ surges obtained 
with DST, and in \S 3.2, compare them with 171{\AA}-ejections.
Then we discuss the magnetic configuration of this region and 
possible mechanisms for the surge activity in the light bridge 
in \S 3.3.
Our results are summarized in \S 4.

\section{OBSERVATIONAL DATA}

We have observed the surge activities along the light bridge in 
the active region NOAA 8971 (N20$^{\circ}$ , W55$^{\circ}$ ) with DST 
from 23:00 UT on 2000 May 01 to 05:30 UT on May 02.
The H$\alpha$ monochromatic images were obtained 
with the Zeiss Lyot filter of 0.25 {\AA} $\;$passband 
and SONY laservideo disc recorder sequentially in 3 wavelengths: 
H$\alpha \pm 0.0$ {\AA}, $+ 0.6$ {\AA}, and $- 0.6$ {\AA}.
The successive wavelength-change and recording were controlled 
with a personal computer, 
and the time capture for each wavelength was 12 seconds.
In this study we mainly used the H$\alpha - 0.6$ {\AA} images, 
in which their ejecting motions are clearly seen.

The EUV images of this region obtained with {\it TRACE} are available 
from 04:30 UT to 06:00 UT on May 02.
They also show conspicuous ejections along the light bridge.
The {\it TRACE} 171 {\AA} images were used to compare 
the features of the hot ejections of about 10$^6$~K 
with those of H$\alpha$ cool surges of about 10$^4$~K.
To co-align the H$\alpha$ images with the EUV images, we used a {\it TRACE} 
1600 {\AA} image.
There are a few soft X-ray images obtained with SXT (Soft X-ray Telescope) 
aboard {\it Yohkoh} \citep{Tsu1991} during the time interval 
from 05:20 UT to 05:30 UT on May 02.
To process the {\it TRACE} and {\it Yohkoh} SXT images, we used the solar 
software of IDL.

In Figure 1, the times of H$\alpha$ and EUV observations are summarized, 
where each time is represented by plus (+) sign.
The second and the fourth rows show the times when the ejections 
along the light bridge are distinctly identified 
at H$\alpha$ and 171 {\AA}, respectively, 
and numbered thick lines in both rows show that 
they are especially ``conspicuous'' ejections. 
We mainly studied the event which occurred at 04:46 UT on May 02 
(\#(7) in Figure 1, 2, and Table 1), because the ejection was clearly seen 
both in H$\alpha$ and in EUV.

\section{RESULTS AND DISCUSSIONS}

\subsection{Motions of H$\alpha$ surges}

During the observation of DST, which was about 6.5 hours, 
H$\alpha$ surges were ejected intermittently from the light bridge 
of the sunspot umbra in NOAA 8971 (see Figure 1).
Figure 2 shows eight surges which extended to larger than 11,000 km in 
apparent length. 
They are the largest among a number of surges observed in this light bridge.
In Figure 2, the {\it top left} is the image at H$\alpha - 5.0$ {\AA}, 
and the others are at H$\alpha - 0.6$ {\AA}.
The mean apparent velocity of these surges is about 40~${\rm km s^{-1}}$, 
the mean apparent maximum length is about 17,000 km, and 
the mean lifetime is about 10 minutes.
These features of the eight surges are listed in Table 1.
We correct these values for the the projection effect.
Assuming that the surges are vertically ejected from the solar surface, 
we get the velocity of 50~${\rm km s^{-1}}$ and the maximum length 
of 20,000 km.
The surges we observed are a few times smaller in velocity 
and about an order of magnitude smaller in maximum length than 
those described by \citet{Tan1995}.

\subsection{Comparison with {\it TRACE} and {\it Yohkoh} SXT images}

We also found ejections from the light bridge in 171 {\AA} images 
obtained with {\it TRACE}. 
These 171{\AA}-ejections have occurred intermittently just as H$\alpha$ 
surges. 
For the event of 04:46 UT (\#(7) in Figure 1, 2, and Table 1), 
we compared the H$\alpha$ surge with the 171{\AA}-ejections 
with respect to their morphological and dynamical characteristics.

Figure 3 shows the evolution of the ejection in H$\alpha - 0.6$ {\AA}
({\it top}) and {\it TRACE} 171 {\AA} ({\it middle; negative}).
The {\it bottom right} panel is a {\it Yohkoh} SXT image 
overlaid with the contour of the sunspot umbra.
The three panels in the {\it bottom left} box show the comparison 
of the appearance and the site of the ejections.
The rightmost panel in the box gives the spatial relation among 
the sunspot umbra, the H$\alpha$ surge, and {\it TRACE} 171 {\AA} loop,
where they are displayed in dark gray, 
light gray, and black curved line, respectively.
The timing and location of the ejections in {\it TRACE} 171 {\AA} images 
are almost the same as those of the surges at H$\alpha -0.6$ {\AA}. 
Furthermore, the velocity of the 171{\AA}-ejection is about 
40~${\rm km s^{-1}}$, and it is nearly equal to that of H$\alpha$ 
surge (see Table 1).
However, the appearance of the ejections are different between 
the H$\alpha$ images and the EUV images; 
that of H$\alpha$ surge is like a jet, 
while that of 171{\AA}-ejection looks like a reverse U-shaped loop.

The H$\alpha$ surges seem to be ejected along some open magnetic field lines 
of about 5,000 km in width, and about 12,000 km in length.
On the other hand, the 171{\AA}-ejections (hot plasma) are reverse U-shaped 
loops that trace the edge of H$\alpha$ surges (cool plasma) 
(see the {\it bottom middle} cartoon in Figure 3).
The separation of two foot points of the loop is about 5,000 km, 
and the loop top is about 12,000 km in hight.
The growth of the reverse U-shaped 171 {\AA} loops indicates the existence 
of some bipolar magnetic polarities in the light bridge of the sunspot umbra.
In addition, as will be discussed below with the magnetograms (\S 3.3), 
the emergence of some new magnetic flux is probably occurring there.
We examined soft X-ray images for the same region obtained with
{\it Yohkoh} SXT, which provide information of the plasma of 
much higher temperature (more than 3 MK).
However, we did not find any SXT ejections according the locations of 
the H$\alpha$ surges and EUV ejection in the light bridge.
Such difference between the appearance of cool (about $10^4$~K) surges 
in H$\alpha$ and hot plasma ejections in 171 {\AA} 
(about $10^6$~K) and in soft X-ray (more than 3 MK) images 
indicate some dynamical and thermal characteristics of plasma ejection 
in the light bridge of the sunspot umbra.
They should be explained by a proper model of the accelerating 
and heating mechanism.

\subsection{Magnetic Configuration}

The continuous surge activity, which was found in the light bridge 
of the sunspot umbra (\S 3.1), is considered to be evidence of 
emerging magnetic flux (e.g. Kurokawa \& Kawai 1993).
The reverse U-shaped loop seen in {\it TRACE} images (\S 3.2) also suggests 
the emergence of a bipolar magnetic flux.

Figure 4 shows the magnetograms on April 27 ({\it top}; near the disk center) 
and on May 02 ({\it bottom}; near the northwest limb, 
N20$^{\circ}$, W55$^{\circ}$) obtained with 
the Michelson Doppler Interferometer (MDI) on board the Solar and 
Heliospheric Observatory ({\it SOHO}) \citep{Sch1995}.
The sunspot has negative polarity (black). 
The polarity of the light bridge is also negative, though its field strength 
is much weaker than the main sunspot umbra (April 27 {\it top}).
On May 02 ({\it bottom}), positive polarity (white) is seen at the location 
on the light bridge.

Since the sunspot is located close to the solar limb 
(55$^{\circ}$) on May 02, 
it is difficult to determine whether this opposite polarity is 
an indication of the newly emerging magnetic flux or whether 
it appears only due to the projection effect of the negative polarity field.
We cannot exclude the possibility that some negative polarity 
which is inclined more than about 40 degrees from the normal 
produces the fake positive polarity.
Nevertheless, we suggest that unresolved and small newly emerging magnetic 
flux plays an essential role in the long-lasting surge activity.
To conclude that new bipolar magnetic fluxes really emerge 
in light bridges and produce such surge activities 
by magnetic reconnection, 
we need more precise observations of magnetic fields of light bridges, 
near the disk center and with higher spatial resolution.

\section{SUMMARY}

We studied dynamical characteristics of H$\alpha$ surges 
along a light bridge in a sunspot umbra with the DST at Hida Observatory.
The surge activities continued intermittently for a long time, 
at least 6.5 hours.
The apparent velocity of the surges was about 50 ${\rm km s^{-1}}$ 
in average, and it is typical for that of an H$\alpha$ surge.
We also studied the ejections from the light bridge observed 
in EUV coronal line of 171 {\AA} with {\it TRACE} for the first time.
The apparent velocities of 171{\AA}-ejections were almost equal to 
those of H$\alpha$ surges.
We could not find any ejections from the light bridge in {\it Yohkoh} 
SXT images.
This means that the temperature of plasma ejected from the light bridge 
is lower than a few MK.
We found morphological differences between H$\alpha$ surges 
and 171{\AA}-ejections; 
the 171{\AA}-ejections seem to be a loop, while the H$\alpha$ 
surges are like jets. 
Examining the magnetogram obtained with {\it SOHO} MDI, 
we suggest that the emergence of new flux occurs in the region, 
though we cannot exclude the possibility that the opposite polarity 
is due to a projection effect.
We need more precise observations of magnetic fields of light bridges.

\acknowledgments

We first acknowledge an anonymous referee for his/her useful comments and 
suggestions that improved this paper much.
We wish to thank Drs. K. Shibata and R. Kitai for fruitful discussions.
We also thank Drs. T. T. Takeuchi, T. J. Wang, K. Yoshimura, H. Kozu, 
T. Morimoto, and H. Isobe for their helpful discussions and comments.
We also thank all the members of Hida observatory for their support
during our observation.
We made extensive use of {\it TRACE} Data Center, {\it SOHO} MDI Data Service, 
and {\it Yohkoh} Data Center.

\clearpage

\begin{table}
\begin{center}
Table 1: Apparent velocity and 
max. length of large surges

\begin{tabular}{rccc}\tableline\tableline
number & time & apparent & apparent \\
       &      & velocity & max. length \\
\# & [UT] & [km s$^{-1}$] & [Mm] \\ \tableline
H$\alpha$ $\;\;\;$(1) & 00:40 & 37.1 & 19.9 \\ 
(2) & 02:38 & 32.3 & 23.2 \\
(3) & 02:50 & 41.0 & 15.9 \\
(4) & 03:19 & 57.3 & 17.0 \\
(5) & 03:57 & 44.0 & 14.2 \\
(6) & 04:27 & 28.3 & 14.2 \\
(7) & 04:46 & 40.4 & 13.9 \\
(8) & 05:32 &      & 15.6 \\ 
mean value & & 38.5 & 16.7 \\ \tableline
EUV $\;\;\;\;\;\;\;\;$ \\
171 {\AA} (7) & 04:46 & 41.5 & \\ \tableline

\end{tabular}
\end{center}
\end{table}

\clearpage

\begin{figure}
\plotone{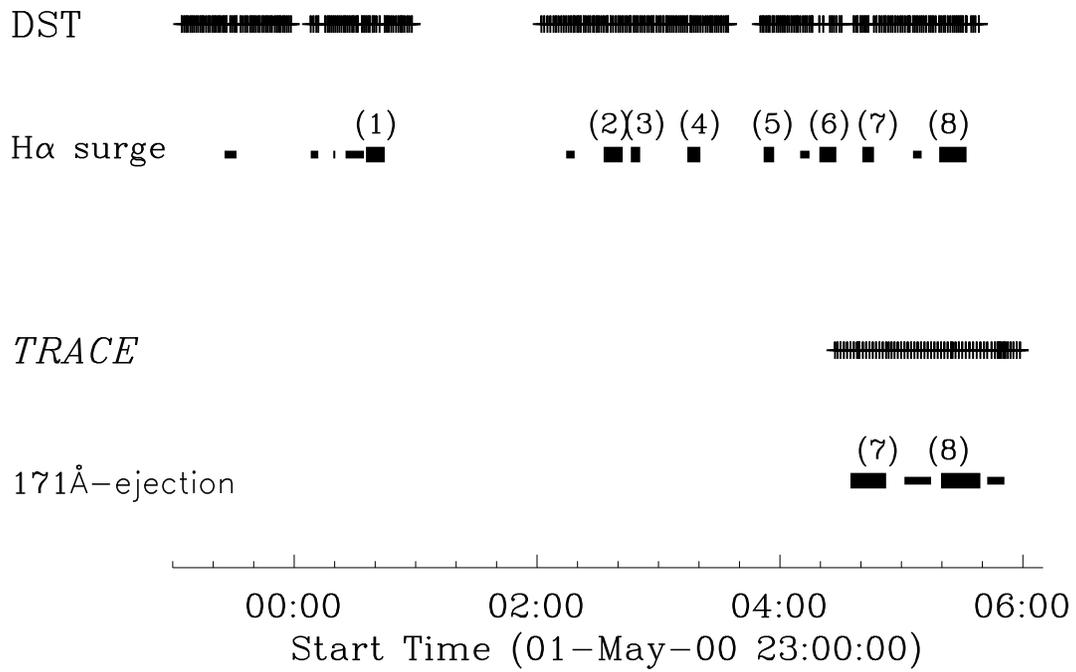}
\caption{
Schematic description of the observation log.
Each observed time is represented by plus (+) sign.
The second and the fourth rows show the ejections seen in 
the H$\alpha$ images and in the EUV images, respectively.
Ejections in both the H$\alpha$ and 171 {\AA} occur recurrently.
\label{fig1}}
\end{figure}

\begin{figure}
\plotone{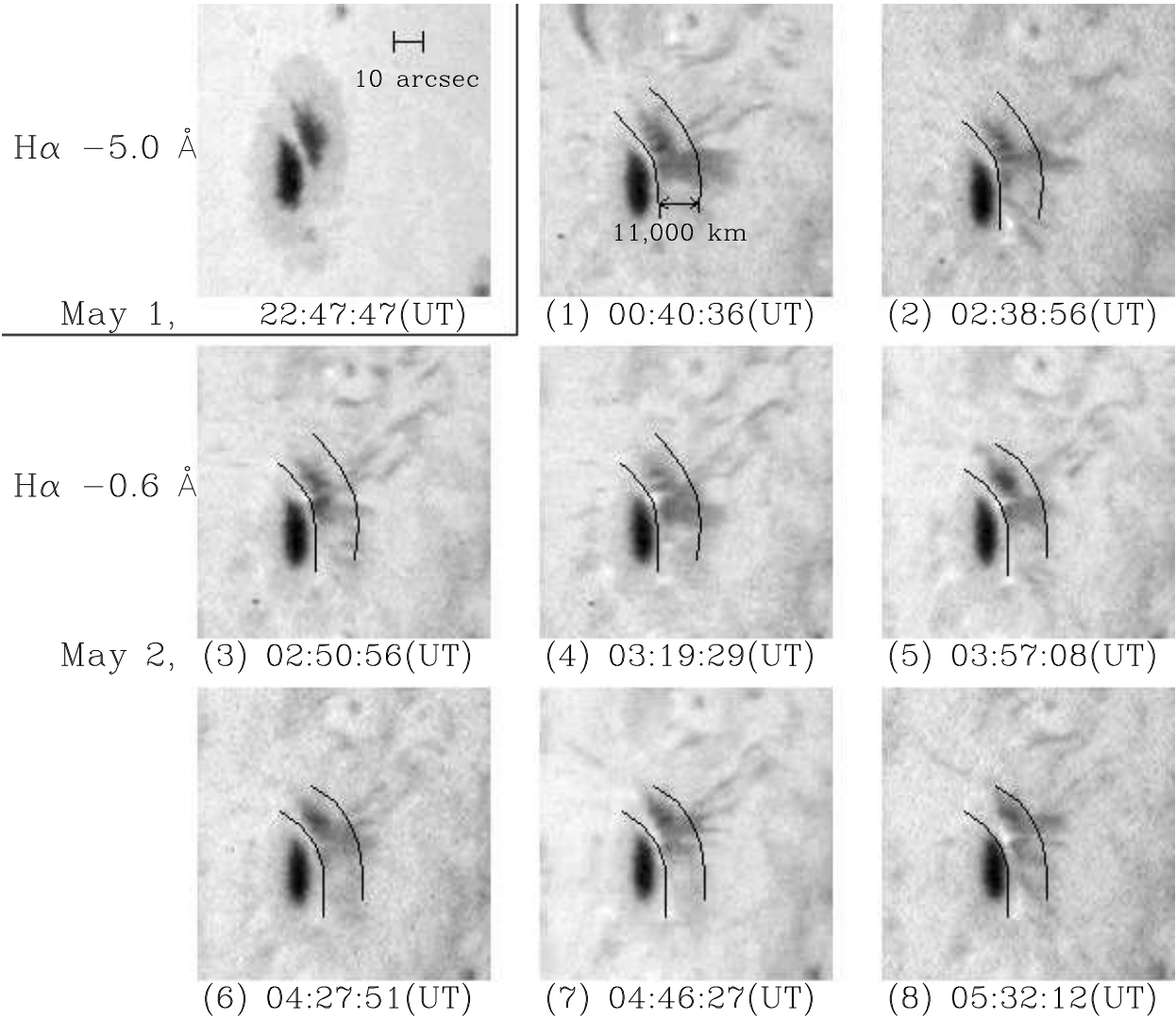}
\caption{
Eight large H$\alpha$ surges.
The celestial north is up, and west is to right.
The {\it top left} image, which clearly shows the light bridge 
studied in this work, 
is the image at H$\alpha -5.0$ {\AA}, and the others are at $-0.6$ {\AA}.
The separation of two curved lines, one of which is along the light bridge, 
is about 11,000~km.
\label{fig2}}
\end{figure}

\begin{figure}
\plotone{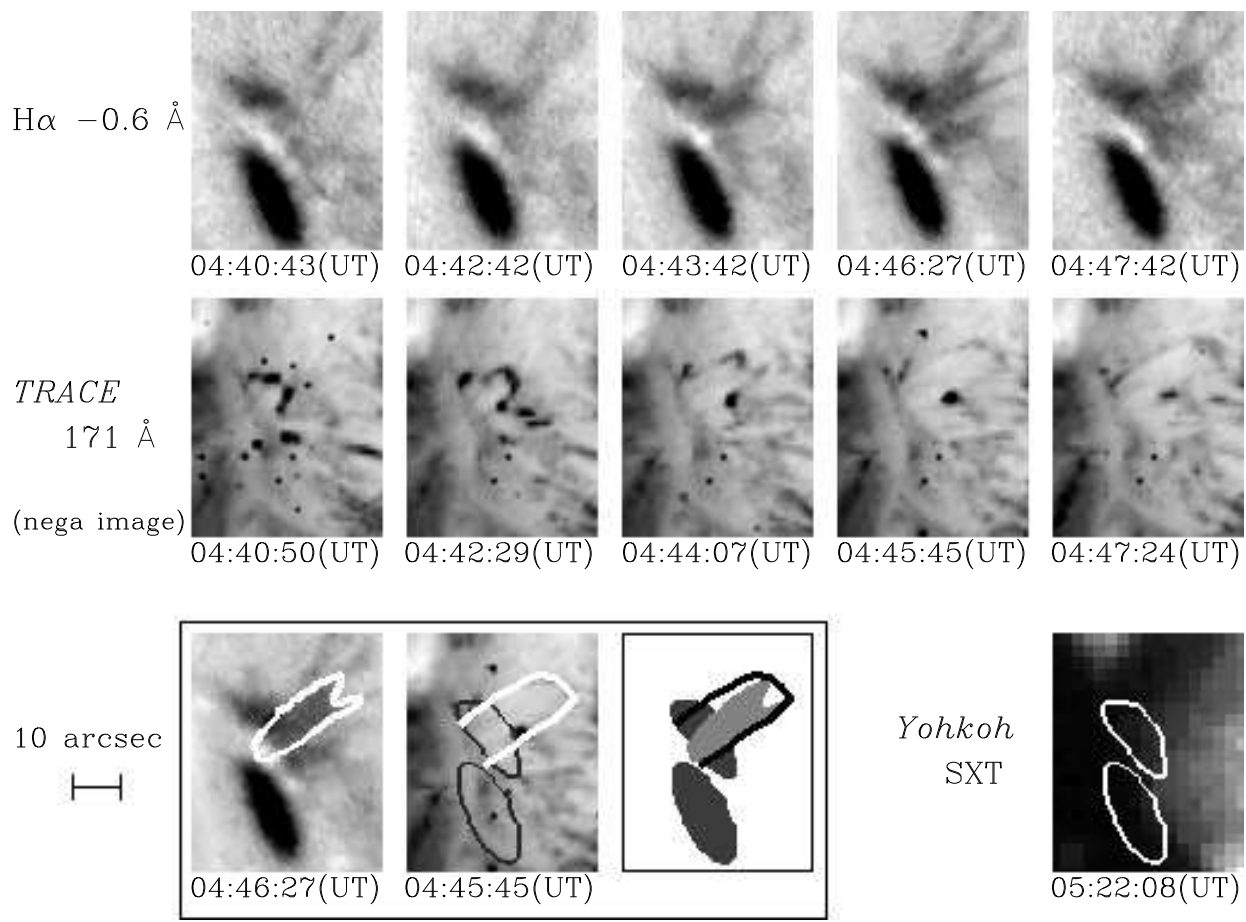}
\caption{
Temporal evolution of \#(7) ejection in DST H$\alpha$ ({\it top}), 
{\it TRACE} 171 {\AA} ({\it middle; negative}).
The solar north is up, and west is to the right.
The {\it bottom right} panel is a {\it Yohkoh} SXT image, 
where two white ellipses show the locations of the sunspot umbrae.
No bright ejection is seen along the light bridge in the 
{\it Yohkoh} SXT image.
The three panels in the {\it bottom left} box give the comparison 
of the appearance and the position of the H$\alpha$ surges 
with those of 171{\AA}-ejections.
In the rightmost panel, the umbra, the H$\alpha$ surge, and {\it TRACE} 
171 {\AA} loop are displayed in dark gray, 
light gray, and black curved line, respectively.
The ejection is seen as a jet in H$\alpha -0.6$ {\AA}, but a loop 
in {\it TRACE} 171 {\AA}.
\label{fig3}}
\end{figure}

\begin{figure}
\plotone{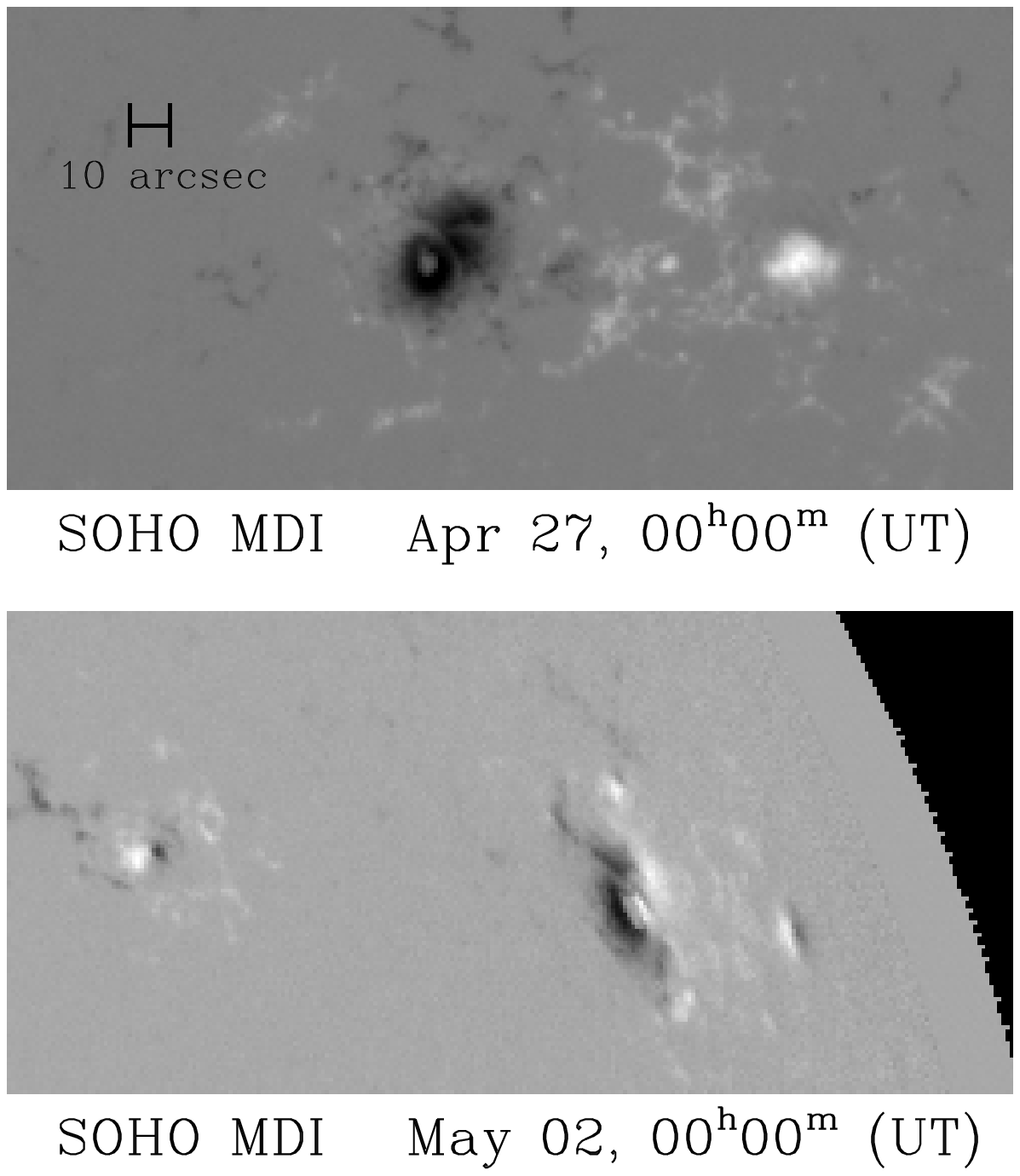}
\caption{
Magnetogram of NOAA 8971 obtained with {\it SOHO} MDI on April 27 
({\it top}) and May 02 ({\it bottom}).
The solar north is up, and west is to right.
The umbrae and the light bridge studied here clearly show a black polarity 
on April 27.
On May 27, however, some fairly white polarity can be seen at the location 
of the light bridge.
\label{fig4}}
\end{figure}

\end{document}